\documentclass[twocolumn,showpacs,preprintnumbers,amsmath,amssymb]{revtex4}
\usepackage{times}
\usepackage{graphicx}
\usepackage{bm}

\newcommand{\half}{\mbox{\small $1 \over 2$}}

\newcommand{\bJ}{{{J}}}
\newcommand{\f}{\frac}
\newcommand{\p}{\partial}
\newcommand{\nn}{\nonumber}
\newcommand{\bea}{\begin{eqnarray}}
\newcommand{\eea}{\end{eqnarray}}
\def\a{\alpha}

\def\d{\delta}

\def\p{\partial} 
\def\nn{\nonumber}

\def\la{\langle}
\def\ra{\rangle}
\def\e{\epsilon}

\def\g{\gamma}

\def\bl{{\bf l}}

\def\bx{{\bf x}}
\def\bv{{\bf v}}

\def\hL{\hat{L}}
\begin{document}
\title{Green-Kubo formula for heat conduction in open systems}
\author{Anupam Kundu$^1$, Abhishek Dhar$^1$ and Onuttom Narayan$^2$}
\affiliation{$^1$ Raman Research Institute, Bangalore 560080, India}
\affiliation{$^2$ Department of Physics, University of California, Santa Cruz, CA 95064}
\date{\today}
\begin{abstract}
We obtain an exact  Green-Kubo type linear response result for the heat
current in an open system.
The result is derived for classical Hamiltonian systems
coupled to heat baths.  Both lattice models and fluid systems are
studied and several commonly used  implementations of heat baths,
stochastic as well as deterministic, are considered. The results are valid in 
arbitrary dimensions and for any system sizes.   
Our results are useful for obtaining the linear response transport properties of
mesoscopic systems. Also we point out that for systems with anomalous heat transport,
as is the case in low-dimensional systems, 
the use of the standard Green-Kubo formula is problematic and 
the open system formula should be used.
\end{abstract}

\maketitle

The Green-Kubo formula~\cite{green54,kubo57b} is a cornerstone of the study of
transport phenomena. For a system governed by Hamiltonian dynamics, the
currents that flow in response to small applied fields can be related
to the equilibrium correlation functions of the currents.
For the case of heat transport the Green-Kubo formula (in the classical
limit, which this paper is restricted to) gives:
\begin{eqnarray}
\kappa &=& \lim_{\tau\rightarrow\infty}\lim_{L\rightarrow\infty} 
\frac{1}{k_BT^2L^d}
\int_0^\tau dt \langle J(t) J(0)\rangle ~,
\label{kubo_n}
\end{eqnarray}
where $\kappa$ is the thermal conductivity of a $d$-dimensional system
of linear dimension $L$ at temperature $T$.
The autocorrelation function on the right hand side is evaluated in
equilibrium, without a temperature gradient. The total heat current in the system is
$J(t) = \int j({\bf x}, t) d{\bf x}$, where $j(\bx,t)$ as the  heat
flux density. 
The order of the limits in Eq.~(\ref{kubo_n}) is important. 
With the correct order of limits, one can calculate the correlation
functions with arbitrary boundary conditions and apply Eq.~(\ref{kubo_n}) to obtain the
response of an open system with reservoirs at the ends.
There have been a number of derivations of Eq.~(\ref{kubo_n}) 
by various authors \cite{green54,kubo57b,others}.

There are several situations where the Green-Kubo formula in
Eq.~(\ref{kubo_n}) is not applicable. 
For example, for the small structures that are studied in mesoscopic
physics, the thermodynamic limit is meaningless, and one is interested
in the conductance of a specific finite system. 
Secondly, in  many low dimensional systems, heat transport is anomalous
and the thermal conductivity diverges \cite{lepri03}.
In such cases it is impossible to take the limits as
in Eq.~(\ref{kubo_n}); one is there interested in the thermal
conductance as a function of $L$ instead of an $L$-independent thermal
conductivity. 
The usual procedure that has been followed in the heat conduction
literature is to put a cut-off at $t_c \sim L$, in the  upper limit in
the Green-Kubo integral \cite{lepri03}. 
There is no rigorous justification of this assumption.
A  related case is that of integrable systems, where the infinite time
limit of the correlation function in Eq.~(\ref{kubo_n}) is non-zero.
Another  way of using the Green-Kubo formula for finite systems
is to include the infinite reservoirs also while applying
the formula and this was done,
for example, by Allen and Ford \cite{allen68} for heat transport and
by Fisher and Lee \cite{fisher81} for electron transport. 
Both these cases are for non-interacting systems and the final expression
for conductance is what one
also obtains from the nonequilibrium Green's function approach, a
formalism of transport commonly used in the mesoscopic literature.
More recently, it has been shown that
Green-Kubo like expressions for finite  open systems can be
derived rigorously by using the steady state fluctuation theorem (SSFT)
\cite{gallavotti96a,lebowitz99,bellet02,gaspard07}.

In this paper, we derive a Green-Kubo like formula for open systems,
without invoking the SSFT. Our  proof applies to all classical systems, of arbitrary size and
dimensionality, with a variety of commonly used implementations of heat
baths. The proof consists in first solving the equation of
motion for the phase space probability distribution to find the
${\cal{O}}( \Delta T)$ correction to the equilibrium distribution
function. The average current at this order can then be expressed in
terms of the equilibrium correlation $\la \bJ(t) J_{fp}(0) \ra $,
where $J_{fp}$ is a specified current operator. 
Secondly we use the energy continuity equations to relate 
two different current-current correlation functions, namely $\la
\bJ(0) \bJ(t) \ra$ and $ \la \bJ(0) J_b(t) \ra$ where $J_b$ is
an instantaneous current operator involving heat flux from the baths.
Finally one relates $\la \bJ(0) J_b(t) \ra $ to $\la \bJ(0) J_{fp}(t) \ra$
and then, using time-reversal invariance, to $\la \bJ(t) J_{fp}(0)
\ra$. For baths with  stochastic dynamics, time-reversal invariance
follows from the detailed balance principle, which is an essential
requirement of our proof.

We first give a proof of our linear response result for a $1D$ lattice model
with white noise Langevin baths. 
We consider the following general Hamiltonian:
\begin{eqnarray}
H= \sum_{l=1}^N \left[~\f{m_l{v}_l^2}{2}+V(x_l)~\right] + \sum_{l=1}^{N-1} U(x_l-x_{l+1})~, 
\label{ham1d}              
\end{eqnarray} 
where $\bx=\{ {x}_l\}$, $\bv=\{v_l\}$ with  $l=1,2...N$ denotes displacements of the
particles about their equilibrium positions  and their velocities, 
and $\{m_l\}$  denotes their  masses.
The particles at the ends are connected to 
two white noise heat baths of temperatures $T_L$ and $T_R$ respectively. 
The equations of motion of the system are given by:
\begin{eqnarray}
m_l\dot{v}_l &=& f_l -\d_{l,1} [ {\gamma^L}{v}_1 - {\eta}^L] 
-\d_{l,N} [ {\gamma^R}v_N - {\eta}^R~], \label{eqmlang}              
\end{eqnarray}
where $f_l= -\p H/\p{x_l}$, 
and  ${\eta}^{L,R}(t)$  are Gaussian noise terms with zero mean and
satisfying the fluctuation dissipation relations:   
$\la {\eta}^{L,R}(t){\eta}^{L,R}(t')\ra = 2{\gamma}^{L,R}{k}_B{T}_{L,R}\delta(t-t')~.$

In the first part of the proof we obtain an expression for the nonequilibrium steady state
average $\la \bJ \ra_{\Delta T}$, at linear order in $\Delta T$, and then
we will relate this to the equilibrium correlation function $\la \bJ(t)
\bJ(0) \ra$ \cite{fnote}. 
Corresponding to the stochastic Langevin equations in Eq.~(\ref{eqmlang}), one
has a Fokker-Planck (FP)equation for the phase space distribution
$P(\bx,\bv,t)$. Setting $T_L=T+\Delta T/2$ and $T_R=T-\Delta T/2$  we
write the FP equation in  the following form:

\vspace{-0.05cm} 
\bea
&& \f{\p P(\bx,\bv,t)}{\p t}=  \hL P(\bx,\bv,t) + \hL^{\Delta T}
P(\bx,\bv,t)~, \label{fpeqn} \\
&& {\rm where}~~~\hL(\bx,\bv) = \hL^H 
 +  \sum_{l=1,N} \f{\gamma^l}{m_l} \f{\p}{\p v_l} \left( v_l+\f{k_B T}{m_l} \f{\p}{\p
  v_l}\right) \nn \\ 
&& \hL^{\Delta T}(\bv) =
\frac{k_B \Delta T}{2}~\left(~\f{\gamma^L}{m_1^2} \frac{\partial^2}{\partial v_1^2} 
- \f{\gamma^R}{m_N^2}\frac{\partial^2}{\partial v_N^2}~\right) ~,
\eea

\vspace{-0.05cm}
where $\hL^H=-\sum_l [~v_l~{\p}/{\p x_l} +(f_l/m_l)~{\p}/{\p v_l}~]$ is the
Hamiltonian Liouville operator and $\gamma^1=\gamma^L, \gamma^N=\gamma^R$.    
For $\Delta T=0$ the steady state 
solution of the FP equation is known and is just the usual equilibrium
Boltzmann distribution $P_{0}=e^{-\beta H}/Z$, where $Z=\int d \bx d
\bv e^{-\beta H}$ is the canonical partition function [$\beta=(k_B
  T)^{-1}$]. It is easily verified that $\hL P_{0}=0$. 
For $\Delta T \neq 0$, we solve Eq.~(\ref{fpeqn}) by perturbation
theory, starting from the equilibrium solution at time $t=
-\infty$. Writing $P(\bx,\bv,t) =P_{0}+p(\bx,\bv,t)$, we obtain   
the following  solution at ${\cal{O}}(\Delta T)$:
\begin{eqnarray}
&&p(\bx,\bv,t) =  \int_{-\infty}^{t}~dt'~e^{\hL(t-t')}
~\hL^{\Delta T}~P_{0}(\bx,\bv)~ \nn \\
&&= {\Delta \beta} \int_{-\infty}^{t}~dt'~e^{\hL (t-t')}
~J_{fp}(\bv)~P_{0}(\bx,\bv)~, \nn \\
&&{\rm with}~J_{fp}(\bv)= {(\Delta \beta~P_0)}^{-1}
  \hL^{\Delta T} P_0  \nn \\
&&=-\f{\gamma^L}{2} \left[ v_1^2 - \frac{k_BT}{m_1}\right ]
+\f{\gamma^R}{2}\left[ v_N^2 - \frac{k_BT}{m_N}\right]~.~~~~~  
\end{eqnarray}

To define the current operator, one first defines the local energy density
at the $l$'th site: $\epsilon_l=m_l
v_l^2/2+V(x_l)+\f{1}{2}[U(x_{l-1}-x_l)+U(x_{l}-x_{l+1})]$. Taking a time
derivative gives the energy continuity equation 
\bea
d \e_l/d t+j_{l+1,l}-j_{l,l-1}&=&j_{1,L}~\d_{l,1}+j_{N,R}~\d_{l,N}~, \label{conteq}\\
{\rm where}~~~~ j_{l+1,l} &=& \f{1}{2} (v_{l}+v_{l+1}) f_{l+1,l} \nn
\eea 
gives the current from the $l$'th to the $l+1$'th site ($f_{l+1,l}$ is the 
force on $l+1$'th particle due to $l$'th particle). 
We define the total current flowing through the system as 
$\bJ= \sum_{l=1}^{N-1} j_{l+1,l}~$. 
The expectation value  of the total current is then given by:
\begin{eqnarray}
\la \bJ \ra_{\Delta T}&=& \int d\bx d \bv~ \bJ ~p(\bx, \bv) \nn \\
&=&{\Delta \beta}~\int_{0}^{\infty}dt~
\int d\bx d \bv~ \bJ~ e^{\hL t}~ J_{fp}~ P_{0} \nn \\ 
&=&{\Delta \beta}~\int_{0}^{\infty}dt~
 \la \bJ(t) J_{fp}(0) \ra~. \label{reln1}
\end{eqnarray}

There are two parts of the proof remaining. Let us define 
the current variable $J_b$ as the mean of the instantaneous heat currents flowing
into the system from the left reservoir 
and flowing out of the system to the right reservoir. Thus we have 
\bea
J_b(t) &=& \f{1}{2}( j_{1,L}-j_{N,R} ) \label{jeta} \\
{\rm where} ~~~{j}_{1,L}(t)&=& - {\gamma}^L {v}_1^2(t)+ {\eta}^L(t){v}_1(t)~,  \nn \\
{j}_{N,R}(t)&=&- {\gamma}^R{v}_N^2(t)+  {\eta}^R (t){v}_N(t) ~.
\end{eqnarray}
The two remaining steps then consist of proving the relations:
\bea
\la \bJ(0) J_b(t) \ra~&=&\la \bJ(0) J_{fp}(t) \ra   \nn \\
 &=& -\la \bJ(t) J_{fp}(0) \ra, \label{reln2}\\
{\rm and}~ \int_0^\infty dt \la \bJ(t) \bJ(0) \ra &=& {(N-1)} \int_0^\infty dt \la
\bJ(0) J_b (t) \ra~ .~~~~~~~ \label{reln3}
\eea
The first line in Eq.~(\ref{reln2}) follows from $\langle J(0)\rangle = 0$
and the result:
\bea 
\la \bJ(0) \eta^L(t) v_1(t) \ra= \la \bJ(0) \eta^R (t) v_N(t) \ra=0~, \label{noisereln}
\eea
which  can be proved by making use of  Novikov's theorem \cite{novikov65,fnote1}. 
The second line in Eq.~(\ref{reln2}) is a statement of time-reversal
symmetry. To prove this we write
$\la J_{fp}(t) J(0) \ra = \int dq \int dq' J_{fp}(q)
J(q') P_0 (q') W(q,t|q',0)$   
where $W(q,t|q',0)$ denotes the transition probability 
from $q'=(\bx',\bv')$  to  $q=(\bx,\bv)$ in time $t$.
Then, using the detailed balance principle
$W(\bx,\bv,t|\bx',\bv',0)P_0(\bx',\bv')=W(\bx',-\bv',t|\bx,-\bv,0)P_0(\bx,-\bv)$
(see \cite{haken,risken,kurchan}) and the fact that $J$ is odd in the velocities while $J_{fp}$
is even, one gets $\la J_{fp}(t) J(0) \ra= -\la J_{fp}(0) J(t) \ra$. A
more direct but equivalent proof is given in [\onlinecite{fnote2}].

We next prove the relation given by Eq.~(\ref{reln3}). 
Let us define  $D_l(t)=\sum_{k=1}^l \e_k -\sum_{k=l+1}^N \e_k$ for
$l=1,2,...N-1$.
Then from the continuity equation Eq.~(\ref{conteq}) 
 one can show that
\bea
d{D_l}/dt=  -2 j_{l+1,l}(t) +2 J_b(t)~. \label{ddoteq}
\eea 
We multiply this equation by $ J(0)$, take a steady state
average, and integrate over time from $t=0$ to $\infty$.  
Since $D_l J $ has an odd power of velocity we therefore get $\la D_l(0) J(0) \ra=0$. 
Also $\la D_l(\infty) J(0) \ra =\la D_l(\infty)\ra ~\la J(0) \ra= 0$ and
using these  we immediately get $\int _0^\infty dt
\la j_{l+1,l}(t) J(0) \ra =  \int_0^\infty dt \la 
J_b (t) J(0) \ra$. Summing over all bonds 
thus proves Eq.~(\ref{reln3}). With a temperature difference $\Delta
T$ between the reservoirs, the steady state current between the
reservoirs and  
the system $\langle I\rangle_{\Delta T}$ is equal to $\la \overline j
\ra$ where $\overline j = J/(N-1).$
Using Eqs.~(\ref{reln1},\ref{reln2},\ref{reln3}), the conductance is 
given by:
\begin{eqnarray}
G = \lim_{\Delta T\rightarrow 0}\frac{\langle \overline j\rangle_{\Delta T}}{\Delta T} =
\frac{1}{k_BT^2}\int_{0}^{\infty}dt \la \overline j(t) \overline j(0)\ra~~, 
\label{LROP}
\end{eqnarray}
which is the central result of the paper. 

The above proof can be extended to the case where the noise from the
baths is exponentially correlated in time \cite{kundulong}. 
Here we will outline the proof for two other models: a deterministic bath model coupled to a 
lattice Hamiltonian and another model where Maxwell baths are coupled
to a fluid system. 

{\emph{ Nose-Hoover baths}}:  In this case the equations of motion are given by:
$m_l\dot{v}_l = f_l -\d_{l,1} \zeta_L {v}_1 -\d_{l,N} {\zeta_R v_N}$ 
where $\zeta_{L,R}$ are themselves dynamical evolving by the 
equations:
\bea
\dot{\zeta_L}&=& ( {\beta_L m_1 v_1^2}-1 )/\theta_L \nn \\
 \dot{\zeta_R}&=&( {\beta_R m_N v_N^2}-1 )/\theta_R~. \nn 
\label{new_eqmnh}
\eea
For small $\Delta T$, we then write an equation of motion for the extended distribution
function $P(\bx,\bv,\zeta_L,\zeta_R,t)$ and find that this has the same form as
Eq.~(\ref{fpeqn}) but now with: 
\bea
&&\hL^{\Delta T} =\frac{\Delta T}{2k_B T^2}
\left(\frac{m_1 v_1^2}{\theta_L}\frac{\partial}{\partial\zeta_L} 
-\frac{m_N v_N^2}{\theta_R}\frac{\partial}{\partial\zeta_R} \right)~. 
\end{eqnarray}
If $T_L=T_R = T,$ one can verify that the equilibrium phase space
density is given by $\hat P_0 = c~P_0(\bx,\bv)~ \exp [-{\theta_L \zeta_L^2}/{2m_1} -
  {\theta_R \zeta_R^2}/{2m_N}]$~, where $c$ is a normalization
constant (independent of $T_{L,R}$), and we assume convergence to this distribution.
Acting with $\hL^{\Delta T}$ on this, we then obtain:
\begin{equation}
J_{fp}  = \frac{1}{2}
(v_1^2 \zeta_L - v_N^2\zeta_R)~.
\end{equation}
On the other hand, since $-\zeta_L v_1$ is the force from the left
reservoir on the first particle, 
hence $j_{1,L} = -\zeta_L v_1^2$ and similarly, $j_{N,R} =-
\zeta_R v_R^2$. Hence from the definition of $J_b$ in
Eq.~(\ref{jeta}), we obtain $J_{fp}=-J_b$. The rest of the proof is 
similar to the previous case, except that there is a minus sign in the 
right hand side of the first line of Eq.~(\ref{reln2}). 
This minus sign is not reversed in the second line of
Eq.~(\ref{reln2}) since under time reversal $(x,v,\zeta)\rightarrow
(x, -v, -\zeta)$ and therefore both $J$ and $J_{fp}$ change their
signs [see arguments given after Eq.~(\ref{noisereln})].
Hence we finally get the same linear response result of Eq.~(\ref{LROP}).

The generalization to arbitrary dimensions is straightforward and we
outline the white noise case. 
We consider  a 
$d$-dimensional hypercubic lattice with points represented by $\bl = 
(l_1,l_2,...,l_d)$ where $l_{\a}=1,2...N$ with $\a=1,2,..,d$.
Let ${\bx}_{\bl}$ and ${\bv}_{\bl}$  be the $d$-dimensional
displacement  and velocity vectors respectively, of the particle at
$\bl$. Heat conduction is assumed to take place in the  $\a=\nu$ direction because of 
heat baths at temperature $T_L$ and $T_R$ that are attached 
to all lattice points on the two hypersurfaces $l_\nu=1$ and $l_\nu=N$.
The corresponding Langevin equations of motion are:
\begin{eqnarray}
m_\bl \dot{\bv}_\bl= {\bf f}_\bl+\d_{l_\nu,1} [{\bm\eta}^L_{\bl'}-\gamma^L_{\bl'}
  {\bv}_\bl] +\d_{l_\nu,L} [{\bm\eta}^R_{\bl'}-\gamma^L_{\bl'}   {\bv}_{\bl'}] ~,
\label{langevin_d}
\end{eqnarray}
where $\bl=(l_\nu,\bl')$, so that $\bl'$ denotes points on a
constant $l_\nu$ hypersurface.
The noise terms at different lattice points and in different
directions are assumed to be uncorrelated, and
satisfy the usual fluctuation-dissipation relations. 

Defining the layer energy $\e_{l_\nu}=\sum_{\bl'} \e_{\bl}$ and the
interlayer current  $j^\nu_{l_\nu+1,l_\nu}$ we find, following the
same steps as in the $1D$ case, the analogue of Eq.~(\ref{reln3}) with
$J$ replaced by 
$\bJ^\nu= \sum_{l_\nu=1}^{N-1} j^\nu_{l_\nu+1,l_\nu}$ 
and $J_b$ replaced by:
\bea
J^\nu_b&=&(j^\nu_{1L}-j^\nu_{NR})/2 =\f{1}{2} \sum_{\bl'}
\left\{ -\left[ \g^L_{\bl'}  {\bv_{(1,\bl)}^2} -
  {\bm\eta}^L_{\bl'}.{\bv}_{(1,\bl')} \right] \right. \nn \\
&&~~~~~~+ \left. \left[\g^R_{\bl'} {\bv_{(N,\bl')}^2} -
  {\bm\eta}^R_{\bl'}.{\bv}_{(N,\bl')} \right]
 \right\}~.   \label{jetahd}
\eea
Writing the FP equation and acting with $\hL^{\Delta T}$ on
the equilibrium distribution gives:
\vspace{-0.2cm}
\bea
J^\nu_{fp} = 
  \sum_{\bl'} \f{-\g^L_{\bl'}}{2} \left[ {\bv_{(1,\bl)}^2} - 
\frac{dk_BT}{m_{(1,\bl')}} \right] 
+\f{\g^R_{\bl'}}{2} \left[ {\bv_{(N,\bl')}^2} - 
\frac{dk_BT}{m_{(N,\bl')}} \right] ~\nn.
\end{eqnarray}
>From the forms of $J^\nu_b$ and $J^\nu_{fp}$, it is clear that  we
can repeat the arguments for the $1D$ case which led to  Eqs.~(\ref{reln1},\ref{reln2}).
Hence we get Eq.~(\ref{LROP}) with $\overline j$ replaced by
${\overline j}^\nu=J^\nu/(N-1)$.

{\emph{Fluid system coupled to Maxwell baths}}: We first consider a
$1D$ system of particles in a box of length $L$. The end
particles ($1$ and $N$) interact with baths at temperatures $T_L$ and
$T_R$ respectively. Whenever the first
particle hits the left wall it is reflected back with a random
velocity chosen from the distribution: 
$\Pi(v)= m_1~ \beta_L~\theta(v)~v~\exp[-  \beta_L m_1 v^2/2]~$,  
with a similar rule at the right end. Otherwise the dynamics is
Hamiltonian.

We find the FP current by noting that $J_{fp}=(\Delta \beta)^{-1} [\p_t P/P]_{P=P_0}$.
There are two parts to the evolution of the phase space density: the
Hamiltonian dynamics inside the system, and the effect of the heat
baths. 
After a small time interval $\epsilon,$ the phase space density $P({\bf x};
{\bf v}; t + \epsilon)$ is
\begin{eqnarray}
&= \beta_L m_1e^{- \half\beta_L m_1 v_1^2} 
\int_0^\infty\!\!
P(0, {\bf x}^\prime - {\bf v}^\prime \epsilon; 
-v_0, {\bf v}^\prime - {\bf a}^\prime\epsilon; t) v_0 dv_0\nonumber\\
&\qquad\qquad {\rm for}\,\, x_1 < v_1 \epsilon\nonumber\\
&= \beta_R m_N e^{- \half \beta_R m_N v_N^2} 
\int_0^\infty\!\!
P({\bf x}^\prime - {\bf v}^\prime \epsilon, L; 
{\bf v}^\prime - {\bf a}^\prime\epsilon, v_0; t) v_0 dv_0\nonumber\\
&\qquad \qquad {\rm for}\,\, x_N > L + v_N \epsilon\nonumber\\
  &= P({\bf x} - {\bf v}\epsilon, {\bf v} - {\bf a}\epsilon, t) 
\qquad  {\rm otherwise} 
\end{eqnarray}
where the primed variables in the first and second
lines leave out particles $1$ and $N$ respectively. 
(Note that since $0 < x_1$
and $x_N < L,$ the conditions in the second and third lines imply $v_1 >
0$ and $v_N < 0.$)

If $T_L=T_R = T,$ and $P({\bf x}, {\bf v}, t)  = P_0,$ the equilibrium
phase space density for the temperature $T,$ then the phase space density
at time $t+\epsilon$ is the same. Now if $T_{L,R} = T\pm\Delta T/2,$
with $P({\bf x}, {\bf v}, t)$ still equal to $P_0,$ then
\begin{eqnarray}
P({\bf x}; {\bf v}; t + \epsilon) &=& P_0 
+ \frac{\Delta T}{2 T}\bigg[
(\half\beta m_1 v_1^2 - 1) \theta(v_1\epsilon- x_1) \nonumber\\ 
&-&
(\half\beta m_N v_N^2 - 1) \theta(x_N - L - v_N\epsilon)\bigg]P_0.\nonumber
\end{eqnarray}
Dividing by $\epsilon$ throughout and taking $\epsilon\rightarrow 0$
, we see that
\begin{eqnarray}
J_{fp} &=& 
-{1\over 2} (\half m_1 v_1^2 - k_B T) v_1 \delta(x_1) \theta(v_1) \nonumber\\
       &-& {1\over 2}(\half m_N v_N^2 - k_B T) v_N \delta(x_N - L)\theta(-v_N).
\label{jfp_mb}
\end{eqnarray}
We have to use continuum energy density $\e(x,t)$ and current $j(x,t),$ and the total heat
current is now $J = \int j(x) dx$ instead of
$\sum j_{i+1, i}.$ The continuity equation is still valid, and
defining $D(x,t)=\int_0^x dx'\e (x',t)-\int_x^L \e (x',t)$ and
$A(t)=\int_0^L dx D(x)$, we get the analogue of Eq.~(\ref{reln3}): 
\begin{equation}
\int_0^\infty \langle J(t) J(0)\rangle dt = 
L \int_0^\infty \langle J_b(t) J(0)\rangle dt.
\label{cont_mb}
\end{equation}
Here $J_b = \half[j_{1,L} - j_{N,R}]$ as before, and 
\begin{eqnarray}
j_{1,L} &=& \half m_1 v_1 (v_{1,L}^2 - v_1^2)~\delta (x_1)\theta( -v_1)\nonumber\\
j_{N,R} &=& \half m_N v_N ( v_{N, L}^2 - v_L^2)~\delta(x_N -
L)\theta(v_N)~. \nn
\label{jb_mb}
\end{eqnarray}
The $\delta$-functions enforce the condition that the particle is
colliding with the bath, and $v_{1,L}$ and $v_{N,R}$ are the random
velocities with which they emerge from the collision.
Invoking detailed balance, using the explicit forms of $J_{fp}$ and
$J_b$, and the fact that $J(0)$ is uncorrelated with $v_{1,L},
v_{N,R}$ we can show that $\langle J(0) J_b(t)\rangle = - \langle
J(t) J_{fp}(0)\rangle.$ With Eqs.~(\ref{cont_mb}) and (\ref{reln1}),
we obtain Eq.~(\ref{LROP}) with $(N-1)$ replaced with $L.$
The generalization to a $d$-dimensional system is straightforward. First,
any particle can interact with the baths at the ends if it reaches $x=0$
or $x=L.$ 
Including the effect of the components of the velocity transverse to
the heat-flow direction the derivation of Eq.~(\ref{jfp_mb}) gets modified  and gives
\vspace{-0.2cm}
\begin{eqnarray}
J_{fp} &=& 
\sum_l -{1\over 2} (\half m_l {\bf v}_l^2 - \half (d+1) k_B T) v^\nu_l \delta(x^\nu_l) \theta(v^\nu_l) \nonumber\\
       &-& {1\over 2}(\half m_l {\bf v}_l^2 - \half (d+1) k_B T)
v^\nu_l \delta(x^\nu_l - L)\theta(-v^\nu_l). \nn
\end{eqnarray}
\vspace{-0.4cm}

\noindent The expression for $J_b$ changes similarly, so that the final result of
the previous paragraph is still valid.

\vspace{-0.05cm}
{\emph{Conclusions:}}
In this paper we have derived an exact  expression for the linear
response conductance in a system connected to heat baths. Our results are
valid in arbitrary dimensions and have been derived 
both for a solid where particles execute small
displacements about fixed lattice positions as well for a  fluid
system where the motion of particles is unrestricted, and various heat
bath models have been considered. 

The important differences with the usual
Green-Kubo formula are worth noting. In the present formula, one does not need
to first take the limit of infinite system size; the result is
valid for finite systems. 
The fact that a sensible answer is obtained even for a finite system
(unlike the case for the usual Green-Kubo formula) is because here 
we are dealing with an open system.
Secondly the
correlation function here has to be evaluated {\emph{not}} with Hamiltonian
dynamics, but for an open system evolving with heat bath dynamics.  
Finally we note that unlike the usual derivation of the Green-Kubo
formula where the assumption of local thermal equilibrium is crucial,
the present derivation requires no such assumption. The results are
thus valid even for integrable Hamiltonian models, the only
requirement being that they should attain thermal equilibrium when coupled to one or
more heat reservoirs all at the same temperature. 

Our derivation here is based on using both the microscopic equations
of motion and also the equation for the phase space distribution. 
The broad class of systems and heat baths for which we have obtained
our results strongly suggests that they are valid whenever detailed 
balance is satisfied.


\begin{thebibliography}{}
\bibitem{fnote} The $\la...\ra$ denotes a thermal equilibrium
  average. Time-dependent equilibrium correlation functions require an
  averaging both over initial conditions as well as one over paths.   In
  the Fokker-Planck representation this can be obtained 
  using the time-evolution operator, while in the Langevin
  representation, thermal noise occurs explicitly and has to be averaged over.
\bibitem{fnote1} Novikov's theorem: let $\{\eta_i\}$ be a
  set of arbitrary Gaussian noise variables with $\la
  \eta_i(t)\eta_j(t')\ra~=~K_{ij}(t,t')$ and let $H[\eta]$ be a
  functional of the noise variables. Then  
\begin{eqnarray}
<\eta_i(t)H[\eta]>~
=~\sum_j\int \la \eta_i(t)\eta_j(t')\ra~
\Big<\frac{\delta H[\eta]}{\delta \eta_j(t')}\Big>~dt' \nn~, 
\end{eqnarray}
where ${\delta H[\eta]}/{\delta \eta_j(t')}$ represents 
a functional derivative of  $H[\eta]$ with respect to $\eta$.
\bibitem{fnote2} 
An integration by parts followed by the transformation $\bv \to
-\bv$ yields: $\la J(t) J_{fp}(0) \ra= \int d \bx d \bv J 
e^{\hat{L}t} J_{fp} P_0= \int d \bx d \bv J_{fp} P_0 
e^{{\bar\hL} t} J =-\int d \bx d \bv J_{fp} P_0 
e^{\hL^\dagger t} J$ 
where $\hat{T} \bar{\hL}=\hL^\dagger=\hL^H-\sum_{l=1,N}~[v_l-(\beta
m_l)^{-1} \p_{v_l}] (\gamma^l/m_l) \p_{v_l}$ and $\hat{T}$ denotes
time reversal. We now note the  operator
identities $\hL P_0 = P_0 \hL^\dagger$ and consequently $e^{\hL t}
P_0=P_0 e^{\hL^\dagger t}$ which can be proved using the  form of $P_0$. Using this in
the above equation immediately gives: $\la J(t) J_{fp}(0) \ra= -\int d
\bx d \bv J_{fp}  e^{\hL t} J P_0=  -\la J(0) J_{fp}(t) \ra$.
\end{thebibliography}
\end{document}